\documentclass[aps,11pt,aps,amsmath,amssymb,nofootinbib,superscriptaddress]{revtex4-2}
\usepackage{braket}
\usepackage{hyperref}
\usepackage{tikz}
\usepackage{slashed}
\usepackage{blindtext}
\usepackage{float}
\usepackage{physics}
\usepackage{url}
\usepackage[export]{adjustbox}
\usepackage{graphicx}
\usepackage{bbold}

\begin{document}
 \title{Can neutron star discriminate between Dirac and Majorana neutrinos?}
 
 	\author{\small Ashutosh Kumar Alok}
	\email{akalok@iitj.ac.in}
	\affiliation{Indian Institute of Technology Jodhpur, Jodhpur 342037, India}
	
	\author{\small Neetu Raj Singh Chundawat}
	\email{chundawat.1@iitj.ac.in}
	\affiliation{Indian Institute of Technology Jodhpur, Jodhpur 342037, India}
	
	\author{\small Arindam Mandal}
	\email{mandal.3@iitj.ac.in}
	\affiliation{Indian Institute of Technology Jodhpur, Jodhpur 342037, India}
	
	\author{\small Trisha Sarkar}
	\email{sarkar.2@iitj.ac.in}
	\affiliation{Indian Institute of Technology Jodhpur, Jodhpur 342037, India}
	
	 \begin{abstract}
Any observable repercussion of electromagnetic properties of neutrinos will provide a perspicuous signature of new physics.  This includes the  phenomenon of neutrino spin flip in the propinquity of an external magnetic field. In this work, we study the  inklings of spin flip in a neutron star with a radially varying magnetic field and matter density, known as a magnetar, which is also a source of profuse production of neutrinos during its initial stage of thermal evolution. We find that a precise measurement of neutrino flux emerging from a neutron star would reeducate discrimination between Dirac and Majorana neutrinos as the  flux reduction due to spin flip oscillations are different for both types of neutrinos. Further, the reduction of flux is more preeminent  near the surface as compared to the core of a neutron star.

\end{abstract}
\maketitle
\newpage
\section{Introduction}\label{intro}
Ever since Wolfgang Pauli postulated the existence of neutrinos in 1930 as a saviour to the law of conservation of energy and momentum in $\beta$-decay, the inscrutable and ubiquitous neutrinos have played a cardinal role in the understanding of particle physics as well as has an abstruse implications in the domain of astrophysics and cosmology. The phenomenon of neutrino oscillations has established the fact that neutrinos have non-zero and non-degenerate masses \cite{Giunti:2007ry}. This can be considered as the first substantiation in benevolence of physics beyond the Standard Model (SM) of particle physics. However, despite a number of breakthroughs, there are still multifarious unanswered questions in the neutrino physics sector. These include whether neutrinos are Dirac or Majorana particles , what is the absolute scale of mass of neutrinos and $CP$ violation in the leptonic sector. A number of planned experiments\cite{Jamieson:2022uya} in the neutrino sector will chaperone judicious exploration of many of the open questions in neutrino physics as well as scrutinizing new scenarios that are opened up by the recent advances in the domain.

One such facet is neutrino electromagnetic properties. In various theories involving new physics, neutrinos may develop electromagnetic properties via quantum loop effects by radiative corrections and hence can acquire a non zero magnetic moment \cite{Giunti:2014ixa}. Thus, the interaction of neutrinos with any external electromagnetic field can provide an excellent rostrum to elucidate the theories beyond SM. Apart from disquisition of new physics, neutrino electromagnetic properties can also be an exceptional tool to differentiate between the Dirac and Majorana nature of neutrinos. While coupled to a photon corresponding to an electromagnetic field, the form factor of interaction at zero momentum transfer $f_q(q^2=0)$ which can be measured in any experimental set ups, are different for Dirac and Majorana neutrinos which sets a possible distinction between the two kinds. Also, the magnetic moment of Dirac neutrino can be both diagonal and off-diagonal in nature, while Majorana neutrino can only possess transitional magnetic moment \cite{Giunti:2014ixa}.

In the electroweak (EW) theory of SM, neutrinos are massless left handed (LH) Weyl fermions and have zero magnetic moment. In the theories of minimally extended SM (MESM), the RH neutrinos are also taken into account along with the LH correspondent which is capable of providing a Dirac mass to the particle and hence a non zero magnetic moment which is extremely small due to the small mixing between LH and RH current ($\sim10^{-19}\mu_B,~\mu_B$ is the Bohr magneton) \cite{Shrock:1982sc}. Such models are usually left right symmetric (LRSM) and operate under the gauge group of $SU(3)_c \times SU(2)_L \times SU(2)_R \times U(1)_{B-L}$. However, in some of the new physics models involving additional charged scalar particle, the estimated value of neutrino magnetic moment can be quite large ($10^{-12}-10^{-10}\mu_B$) as compared to the value predicted from the MESM theories \cite{Fukugita:1987ti}. The value of the neutrino magnetic moment is constrained from several terrestrial experiments such as GEMMA \cite{Beda:2012zz}, Borexino \cite{Borexino:2017fbd}, XENON1T \cite{Miranda:2020kwy}, XMASS-I \cite{XMASS:2020zke} which relies upon the observation of neutrino-electron scattering. Several astrophysical and cosmological phenomena are also able to put constraints on the magnetic moment \cite{Raffelt:1990pj,Raffelt:1989xu,Raffelt:1992pi}. 

One of the most engrossing reverberations of electromagnetic properties of neutrinos is the phenomenon of spin oscillation occurring in neutrino states in the presence of an external magnetic field. This can have significant consequences in the context of terrestrial detectors corresponding to several astrophysical backgrounds. In fact, the phenomena of neutrino spin flavour precession (SFP) is treated as an alternate theory for the observed abatement in the solar neutrino flux, beside the framework of neutrino flavour oscillation. However, the SFP in magnetic field by itself cannot account for the deficiency in solar neutrino flux. This prospect is ruled out by the KamLAND experiment data. 

In this work, we consider a dense astrophysical compact object having a large magnetic field, such as a neutron star in which we anatomize the phenomenon of SFP and contemplate whether this can mitigate to distinguish between Dirac and Majorana nature of neutrinos. Neutron stars are unique astrophysical objects which are created as a product of stellar evolution of a main sequence star having higher mass and in which matter resides at the extreme environmental conditions. It has a very high intrinsic magnetic field, which can reach even at an extremely high value of $\sim10^{18}$ G at its center. The structure of a neutron star is held by the degenerate pressure of a large number of neutrons present at its core which is also the source of production of neutrinos in copious amounts. Neutrino is the significant heat carrier of the neutron star which cools down the star most efficiently. During the cooling stage of a neutron star, neutrino emission era is initiated after a brief period of thermal relaxation which creates a thermal equilibrium between the core and crust of the star. In the presence of magnetic field, the Dirac neutrinos are converted into its sterile counterpart, whereas the Majorana neutrinos are converted into anti-neutrinos.

The plan of the work is as follows. In the next section, we delineate the formalism of this work. This includes a brief description of the SFP phenomenon followed by a depiction of magnetic field profile in a neutron star. The results are furnished in Sec.~\ref{res}. The conclusions are acquainted in Sec.~\ref{conc}.

\section{Spin flavor oscillations in neutron star}
\label{II}

In this section, we silhouette the formalism of our work. In Sec. \ref{IIA}, we proffer  magnetic moment of neutrino and the phenomena of SFP as its consequence and in Sec. \ref{IIB}, we limn neutron star and its magnetic field profile.
 
\subsection{Neutrino magnetic moment and spin flip}
\label{IIA}

The interaction of neutrino with an external electromagnetic field catalyses four phenomenon: (a) magnetic moment, (b) electric moment, (c) neutrino charge and (d) anapole moment. In this work, we focus on  the magnetic moment of neutrino ($\mu_{\nu}$). In MESM theories, the magnetic moment of Dirac neutrino is diagonal and is given by \cite{Fujikawa:1980yx}
\begin{equation}\label{eq1}
\mu_{ii}^D=\frac{3eG_F m_i}{8\sqrt{2}\pi^2}\approx \frac{3.2\times 10^{-19}m_i}{1 eV} \mu_B.
\end{equation}
The magnetic moment of Majorana neutrino is off-diagonal, as the diagonal counterpart is suppressed due to antisymmetric magnetic moment matrix as required by the CPT invariance. Magnetic moment for Majorana neutrino can be expressed as \cite{Pal:1981rm}
\begin{equation}\label{eq2}
\mu_{ij}^{M}=-\frac{3eG_F}{32\sqrt{2}\pi^2}(m_i \pm m_j)
\sum_{l=e,\mu,\tau} U_{li}^{*} U_{lj} \frac{m_{l}^{2}}{m_{W}^{2}}.
\end{equation}
In eqn. \eqref{eq1} and \eqref{eq2}, $G_F$, $m_i$ and $e$ represent Fermi constant, neutrino mass and the charge of an electron, respectively. Further, $m_l$ and $m_W$ denote mass of the $l$-th lepton and the $W$-boson mass, respectively. Moreover, $U_{ij}$ is the element of the $3\times3$ PMNS mixing matrix. From \eqref{eq1} and \eqref{eq2}, it can be seen that the value of magnetic moment depends on the neutrino mass $i.e.$ larger neutrino magnetic moment, which is necessary to resolve the solar neutrino problem (SNP), requires higher value of neutrino mass which poses a serious issue as the current limit of neutrino mass is extremely small. However, this problem is resolved in several new physics models including ref. \cite{Voloshin:1986ty} in which a charged scalar is added to the SM particle spectrum to provide a large neutrino magnetic moment and at the same time satisfying the current upper bound on the neutrino mass \cite{Fukugita:1987ti}. Apart from this, an additional approximate horizontal symmetry $SU(2)_H$ with the SM gauge group ($SU(3)_c\times SU(2)_L \times U(1)_Y$) is also capable of enhancing the neutrino magnetic moment without altering its current mass limit \cite{Babu:2020ivd}. Non standard interaction (NSI) \cite{Giunti:2014ixa,Healey:2013vka,Papoulias:2015iga,Kharlanov:2020cti} and several $R$-parity breaking supersymmetric (SUSY) \cite{Aboubrahim:2013yfa, Fukuyama:2003uz} models are also capable of providing large magnetic moment of neutrino.

The value of diagonal and transitional magnetic dipole moments are of the order of $10^{-19}\mu_B$ and $10^{-23}\mu_B$ respectively \cite{Babu:2020ivd} as obtained from MESM theories which are far below the reach of experimental facilities. The off-diagonal magnetic moment is much suppressed in comparison to the diagonal magnetic moment because of the GIM mechanism \cite{Giunti:2008ve, Xing:2012gd}. Several reactor experiments such as GEMMA \cite{Beda:2009kx}, TEXONO \cite{TEXONO:2006xds}, ROVNO \cite{Derbin:1993wy} and accelerator experimental set ups such as LAPMF \cite{Allen:1992qe}, LSND \cite{LSND:2001akn} put constraints on the upper bound of neutrino magnetic moment which, on average, lies in the range $10^{-11}-10^{-10}~\mu_B$ through the observation of $\nu-e$ scattering cross section.  Recently an estimation provided by XENON1T experiment shows the transition magnetic moment $\mu_{e\mu}$ to be existing in the range $1.65-3.42\times10^{-11}~\mu_B$ \cite{Babu:2020ivd}. In astrophysics and cosmology from the evolution of stars \cite{Heger:2008er}, plasmon decay in the stellar environment \cite{Borisov:2014cqa}, supernova events \cite{deGouvea:2012hg} and $^4_2 He$ nucleosynthesis \cite{Vassh:2015yza}, it is also possible to impose indirect constraints on the magnetic moment value. The constraint obtained from the red giant branch (RGB) of globular cluster is $\mu_{\nu}<4.5\times 10^{-12}~\mu_B$ \cite{Viaux:2013hca} which is smaller by at least an order than the limits obtained from different terrestrial experiments .

In SM neutrinos are massless electrically neutral Weyl fermions which cannot interact with electromagnetic fields at tree level which leads to zero magnetic moment. However, from higher order perturbative corrections, neutrinos gain non zero magnetic moment via quantum loop effect . The Hamiltonian for the neutrino-photon interaction is given by \cite{Giunti:2014ixa}
\begin{equation}\label{eq3}
\mathcal{H}_{EM}=\bar{\nu}(x)\Lambda^{\mu}\nu(x) A_{\mu},
\end{equation}
where $A_{\mu}$ is the electromagnetic field and $\Lambda_{\mu}$ denotes the vertex fuction containing the electromagnetic properties of neutrino which is a $4\times4$ matrix in the spinor space. For Dirac neutrinos, it is expressed as \cite{Giunti:2014ixa}
\begin{eqnarray}\label{eq4}
\Lambda_{\mu}^D(q) &=& f_Q(q^2)^Dq_{\mu}\gamma_5 -f_M^D(q^2)i\sigma_{\mu\nu}q^{\nu}+f_E^D(q^2)\sigma_{\mu\nu}q^{\nu}\gamma_5 \nonumber\\
&&+f_A^D(q^2)(q^2\gamma_\mu-q_{\mu}\slashed{q})\gamma_5.
\end{eqnarray}
and for Majorana neutrinos, it is given by \cite{Giunti:2014ixa}
\begin{eqnarray}\label{eq5}
\Lambda_{\mu}^M(q) &=& (\gamma_\mu-q_{\mu}\slashed{q}/q^2)[f_{Q}^{M}(q^2)+f_A^M(q^2)q^2\gamma_5] \nonumber\\
&&-i\sigma_{\mu\nu}q^{\nu}[f_M^M(q^2)+if_E^M(q^2)\gamma_5]\,.
\end{eqnarray}
Here $f_i^{D,M}(q^2) (i=Q,M,E,A)$ are the form factors corresponding to the four electromagnetic properties, as mentioned earlier, corresponding to Dirac (D) and Majorana (M) neutrinos, while $q$ denotes the momentum transfer. $\gamma^{\mu}$ denotes Dirac matrices, $\slashed{a}=\gamma^{\mu}a_{\mu}$, $\gamma_5=i\gamma_0 \gamma_1 \gamma_2 \gamma_3$, $\sigma_{\mu\nu}=[\gamma_\mu,\gamma_\nu]$. From eqn. \eqref{eq4} and \eqref{eq5} it is possible to distinguish between Dirac and Majorana neutrinos due to their distinct form factors and consequently electromagnetic vertex factor. It is to be noted further that neutrino electromagnetic properties are invariant under $CP$ transformation for both Dirac and Majorana neutrinos \cite{Broggini:2012df}, although it is violated in the standard $V-A$ theory of weak interaction. The form factors at zero momentum transfer $q^2=0$ are gauge independent and finite which can be measured by the interaction with the external field, $A_{\mu}$ directly at any experimental set up.

One of the most interesting consequences of electromagnetic properties is the precession of neutrino spin in presence of an external magnetic moment which leads to the mixing between LH and RH neutrino current. In the two flavour framework, the Hamiltonian for neutrino state evolution is expressed as $4\times4$ Hermitian matrix. For Dirac neutrino basis $(\nu_{eL},\nu_{\mu L},\nu_{eR},\nu_{\mu R})$, the Hamiltonian is expressed as \cite{Giunti:2014ixa}
\begin{equation}\label{eq6}
H_D=\begin{pmatrix}
\frac{-\Delta m^{2}}{4E_{\nu}}cos 2\theta+V_e & \frac{\Delta m^{2}}{4E_{\nu}}sin 2\theta & \mu_{ee}B_\perp & \mu_{e\mu}B_\perp\\
\frac{\Delta m^{2}}{4E_{\nu}}sin 2\theta & \frac{\Delta m^{2}}{4E_{\nu}}cos 2\theta+V_\mu & \mu_{e\mu}B_\perp & \mu_{\mu\mu}B_\perp \\
\mu^{*}_{ee}B_\perp & \mu^{*}_{\mu e}B_\perp & \frac{-\Delta m^{2}}{4E_{\nu}}cos 2\theta & \frac{\Delta m^{2}}{4E_{\nu}}sin 2\theta \\
\mu^{*}_{e\mu}B_\perp & \mu^{*}_{\mu \mu}B_\perp & \frac{\Delta m^{2}}{4E_{\nu}}sin 2\theta  &  \frac{\Delta m^{2}}{4E_{\nu}}cos 2\theta \\ 
\end{pmatrix}.
\end{equation} 
For Majorana neutrinos in the basis of $(\nu_{eL},\nu_{\mu L},\bar{\nu}_{e},\bar{\nu}_{\mu})$, the evolution matrix is given by \cite{Giunti:2014ixa}
\begin{widetext}
\begin{equation}\label{eq7}
H_M=
\begin{pmatrix}
\frac{-\Delta m^{2}}{4E_{\nu}}cos 2\theta+V_e & \frac{\Delta m^{2}}{4E_{\nu}}sin 2\theta & 0 & \mu_{e\mu}B_\perp\\
\frac{\Delta m^{2}}{4E_{\nu}}sin 2\theta & \frac{\Delta m^{2}}{4E_{\nu}}cos 2\theta+V_\mu & -\mu_{e\mu}B_\perp & 0 \\
0 & -\mu^{*}_{\mu e}B_\perp & \frac{-\Delta m^{2}}{4E_{\nu}}cos 2\theta-V_e & \frac{\Delta m^{2}}{4E_{\nu}}sin 2\theta   \\
\mu^{*}_{e\mu}B_\perp & 0 &  \frac{\Delta m^{2}}{4E_{\nu}}sin 2\theta &  \frac{\Delta m^{2}}{4E_{\nu}}cos 2\theta-V_\mu  \\ 
\end{pmatrix}.
\end{equation}
\end{widetext}
In equation \eqref{eq6} and \eqref{eq7}, $B_\perp$ denote transverse magnetic field, while $\mu$ is the neutrino magnetic moment. Here, $\Delta m^{2}=m_{2}^2-m_{1}^2$. $V_{e}$ and $V_{\mu}$ are the potentials experienced by $\nu_e$ and $\nu_{\mu}$, respectively. Further, $V_e=\sqrt{2}G_F (n_e-n_n/2)$ and $V_\mu=-\sqrt{2}G_F n_n/2$, where $n_e$ and $n_n$ are electron and neutron number density, respectively. In our work, we consider the simplification of zero vacuum mixing $(\theta_{ij}=0)$ which yields the following Hamiltonian reduced to a $2\times2$ matrix, given by \cite{Joshi:2019dcj}
\begin{equation}\label{eq8}
H=\begin{pmatrix}
-\Delta m_{12}^2/4E+\Delta V/2 & \mu_{e\mu}B \\
\mu_{e\mu}B & \Delta m_{12}^2/4E-\Delta V/2
\end{pmatrix}.
\end{equation}
Here $\Delta V=\sqrt{2}G_F \rho Y_e^{eff} /m_N$ with $\rho$ being the matter density. $Y_e$ and $m_n$ denote the effective electron fraction and nucleonic mass, respectively. $Y_e^{eff}=(3Y_e-1)/2$ for Dirac basis, while for  Majorana basis, $Y_e^{eff}=(2Y_e-1)/2$. Following eqn. \eqref{eq8} the neutrino spin flavour evolution equation in presence of an external magnetic field is expressed as \cite{Joshi:2019dcj}
\begin{widetext}
\begin{equation}\label{eq9}
\frac{d^2 \nu_{eL}}{dr^2}-\left(\frac{\mu \dot{B}}{\mu B}+i\zeta \right)\frac{d\nu_{eL}}{dr}+\left[\phi^2+i\frac{d\phi}{dr}+(\mu B)^2-i\phi \frac{\mu \dot{B}}{\mu B}+\phi \zeta\right]\nu_{eL}=0,
\end{equation}
\end{widetext}
where 
\begin{eqnarray}\label{eq10}
\phi&=&\frac{-\Delta m^2}{4E}+\frac{1}{\sqrt{2}}G_F n_e,  \nonumber\\
\zeta&=& \frac{1}{\sqrt{2}}G_F n_n,~ \nu_{eL}\rightarrow \nu_{\mu R},  \nonumber\\
\hspace*{8mm}&=&\sqrt{2}G_F n_n, ~ \nu_{eL}\rightarrow \bar{\nu}_{\mu}\,.
\end{eqnarray}
The transition probability of LH electron neutrino can be obtained after solving \eqref{eq9} alongwith \eqref{eq10}. It is expected that the solution of these equations will generate different neutrino survival probabilities in the case of Dirac and Majorana neutrinos. 

Now, to analyse this phenomena, a very suitable medium with high magnetic field  is required. For this, we have taken into account an astrophysical background such as a neutron star  which is able to satisfy both of these criteria. In the next section, we briefly describe the interior structure and the magnetic field of a neutron star.

\subsection{Neutron star and magnetic field}\label{IIB}
The SFP phenomena has been previously analyzed in several other astrophysical sources \cite{Joshi:2019dcj,Yilmaz:2016ilw,Lichkunov:2020zzx,PhysRevD.104.023018,Adhikary:2022phm,Alok:2022pdn}. We have carried out our analysis of SFP in the matter environment of neutron star due to
\begin{itemize}

\item presence of high magnetic field \cite{Peng:2007uu},

\item extensive production of neutrinos \cite{Yakovlev:2000jp},	

\item there is nearly one billion neutron stars present in our Milky Way galaxy \cite{Chrimes:2021wqi},

\item magnetars are the neutron stars having magnetic field greater than that possessed by any other stellar object \cite{Kaspi:2017fwg}.
\end{itemize}
 All these points make neutron star a very suitable environment for the SFP analysis. 

Neutron stars  are produced as the end stage product of stellar evolution of medium to higher mass stars having mass within the range $1M_{\odot}\leq M \leq 8M_{\odot}$. After complete exhaustion of nuclear fuel at the stellar core, gravitational collapsing of the stellar structure is initiated and the entire core region is packed with iron nuclei because of earlier continuos nucleosynthesis. This leads to  photo disintegration of iron nuclei. After this, via the process of electron capture, a large nuber of neutrons and neutrinos are produced. With further compression of the matter, neutrons start to drip out of the nuclei and with time they build up a degeneracy pressure. At a certain stage the infalling matter from the outer region to the degenerate neutron core is bounced back from it and generates an ultrasonic shock wave which propagates from the core towards the outer region and the entire structure undergoes a violent explosion of supernova with a sudden burst of trapped neutrinos. The stellar structure left out after the supernova forms a proto neutron star in which the hydrostatic equilibrium is set up by the neutron degeneracy pressure. Although the core of the neutron star mainly consists of neutrons, there can still be a small admixture of protons and electrons, possibly with other charged leptons and hyperons. Neutron stars are extremely dense stellar objects with its mass of the order of $2~M_{\odot}$ packed inside a sphere of $10$ km. The interior matter density reaches at a few times of nuclear saturation density. Till now the maximum observed  mass of a neutron star is $2.3~M_{\odot}$ \cite{Menezes:2021jmw,Vidana:2020jhf,Lattimer:2015eaa}. 

After the formation of a neutron star, its interior temperature remains $\sim 10^{10}-10^{11}$ K and there is a temperature gradient existing between the core and the crust. Within the next $10-100$ years, depending on the initial mass of the progenitor star, a thermal relaxation period is started in which a cooling wave passes from the center to the surface of the star to set up a thermal equilibrium inside the star. After this, the neutrino cooling era is initiated and the star starts to cool down very fast with the emission of neutrinos. The neutrino cooling occurs by the set of simultaneous interaction of $\beta$-decay and electron capture by proton, $n\rightarrow p e \bar{\nu}_e, ~ pe\rightarrow n \nu_e$ when the proton fraction in the star is above a certain critical level, following the inequality, $k_{Fp}\leq k_{Fn}+k_{Fe}$ where $k_F$ denotes the wave number of the fermions. However, below this critical point, the same set of reactions occur in presence of a bystander nucleon ($N$) which is less efficient compared to the previous one. The period of neutrino cooling lasts till $\sim 10^5$ years, after which the star loses heat via the emission of photons \cite{Yakovlev:2004yr,Tsuruta:1986qt,Pethick:1991mk}.

As mentioned earlier, a magnetar can possess an extremely high interior magnetic field, even of the order of $10^{18}$ G. However, the origin of magnetic field is still an open question in the domain of astrophysics,although there are several explanation regarding the subject \cite{Spruit:2007bt,Reisenegger:2003pj}. There exist several magnetic field profiles of neutron stars. One of the oldest field profile, which has been useful to many earlier works, is given by \cite{PhysRevLett.79.2176}
\begin{equation}\label{eq11}
B(n_b/n_0)=B_s+B_c[1-exp(-\beta(n_b/n_0)^\gamma)],
\end{equation}
where $B_c$ and $B_s$ represent the magnetic fields at the core and the surface, respectively, $n_0$ is the nuclear saturation density which is fixed for a particular neutron star equation of state (EoS) and $n_b$ is the total baryon number density. $\beta$ and $\gamma$ are the two free parameters of the profile which are set to be random values to obtain the desired magnetic fields at the core and the surface. Recently, a magnetic field profile has been developed which does not contain any free parameter and is expressed as \cite{Chatterjee:2018prm,Chatterjee:2021wsr}
\begin{equation}\label{eq12}
B(x)=B_c (1-1.6x^2-x^4+4.2x^6-2.4x^8).
\end{equation}
Here, $x=r/R$ with $r$ and $R$ being the radial distance from the center of the star and its radius, respectively.
Neutrinos emitted from the neutron star have energy in the MeV energy range \cite{Yakovlev:2000jp} which are possible to be distinguished by the neutrino detectors placed on Earth.

\section{Result and discussion}
\label{res}

In this section we discuss the analysis of neutrino SFP phenomena occurring in the interior of a magnetar with the consideration of both Dirac and Majorana neutrinos and inspect when the magnetar system can possibly show a distinction between these two kinds of scenarios. We consider an isolated magnetar with mass $2.3~M_{\odot}$ and radius of $11.8$ km. The magnetar consists of $npe$ matter and possesses a magnetic field of $10^{18}$ G at the center and $10^{15}$ G at its surface. We consider the universal magnetic field profile given by eqn. \eqref{eq12} inside the star. The matter configuration follows nonlinear Walecka model based on relativistic mean field (RMF) theory containing $\sigma$, $\omega$ and $\rho$-meson \cite{Santos:2004js,Mueller:1996pm} and it obeys the well known GM1 parametrization \cite{Glendenning:1991es}. The expressions for electron ($n_e(r)$) and neutron number density ($n_n(r)$) are obtained from the nonlinear fit, given by
\begin{eqnarray}
\label{eq13}
n_e(r)&=&0.2056-4.502 \times 10^{-4}r-2.307 \times 10^{-3} r^2 -3.026\times 10^{-4} r^3 +8.873 \times 10^{-5} r^4 \nonumber\\
&& -7.668 \times 10^{-6}r^5 +2.415\times 10^{-7}r^6\,, \nonumber\\
n_n(r)&=&0.877+0.01r-0.022\,r^2+5.447\times 10^{-3}r^3-9.613\times 10^{-4}r^4+8.536\times 10^{-5}r^5 \nonumber\\
&& -2.868\times 10^{-6}r^6\,.
\end{eqnarray}
Both the number densities are maximum at the centre of the star and decrease towards the outer regions. In our analysis, we have assumed zero vacuum mixing ($\theta_{12}=0$)  and $\Delta m_{12}^2=7.5\times 10^{-5}$ eV$^2$. The value of the neutrino magnetic moment is assented to be $\mu_{\nu}\sim 10^{-11}\mu_B$.

\begin{figure}
\includegraphics[scale=0.4]{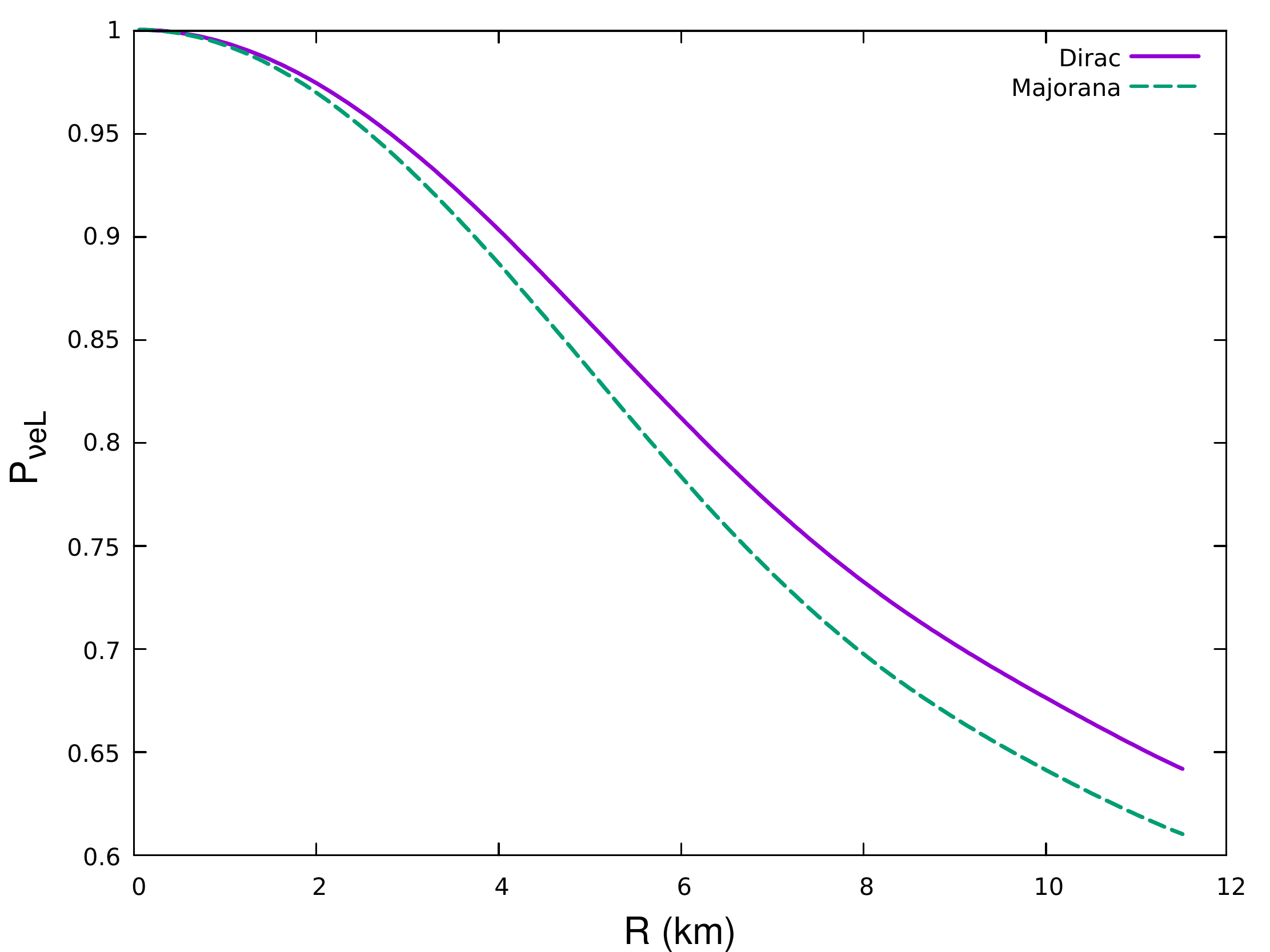}
\caption{Variation of survival spin flip probability ($P_{\nu_{e L}}$) for Dirac (solid line) and Majorana (dashed line) neutrinos with  distance ($R$) from the center of the star. }
\label{f1}
\end{figure}

Fig. 1 brandishes the clear distinction between Dirac and Majorana neutrinos near the surface region of the neutron star, around $R \sim 11.5$ km. We consider an initial flux of $\nu_{eL}$ generated at the inner core of the magnetar and estimate its survival probability at different radial distances ($R$). It can be espied from the plot that the difference between the two kinds of neutrinos is not conspicuous near the inner core region where the matter is ultra dense and resides at the most extreme condition, while the distinction is maximized at the outer layers of the star. On the surface of the star, the survival probability for Dirac and Majorana neutrino are observed to be $\sim 0.64$ and $\sim 0.61$, respectively. Our result implies that the predicted reduction in the flux of Majorana neutrino is larger by $5\%$ as compared to the Dirac neutrinos. In case of Majorana neutrinos, the flux reduction takes place due to its conversion to anti-neutrinos. It is to be noted here that to distinguish between these two pictures, a powerful neutrino detector with high efficiency is imperative.

\begin{figure}
\includegraphics[scale=0.4]{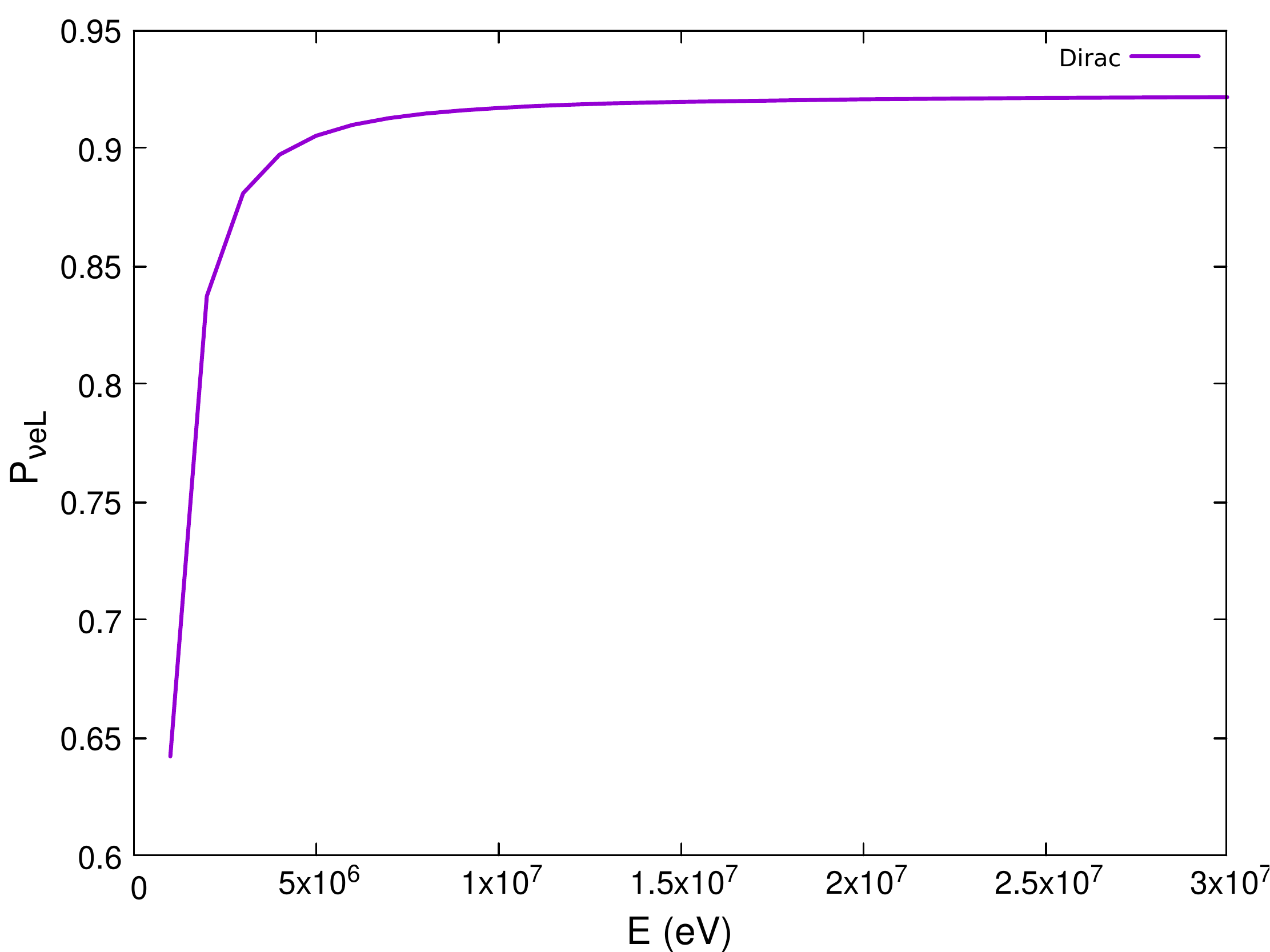}\hspace*{1mm}\includegraphics[scale=0.4]{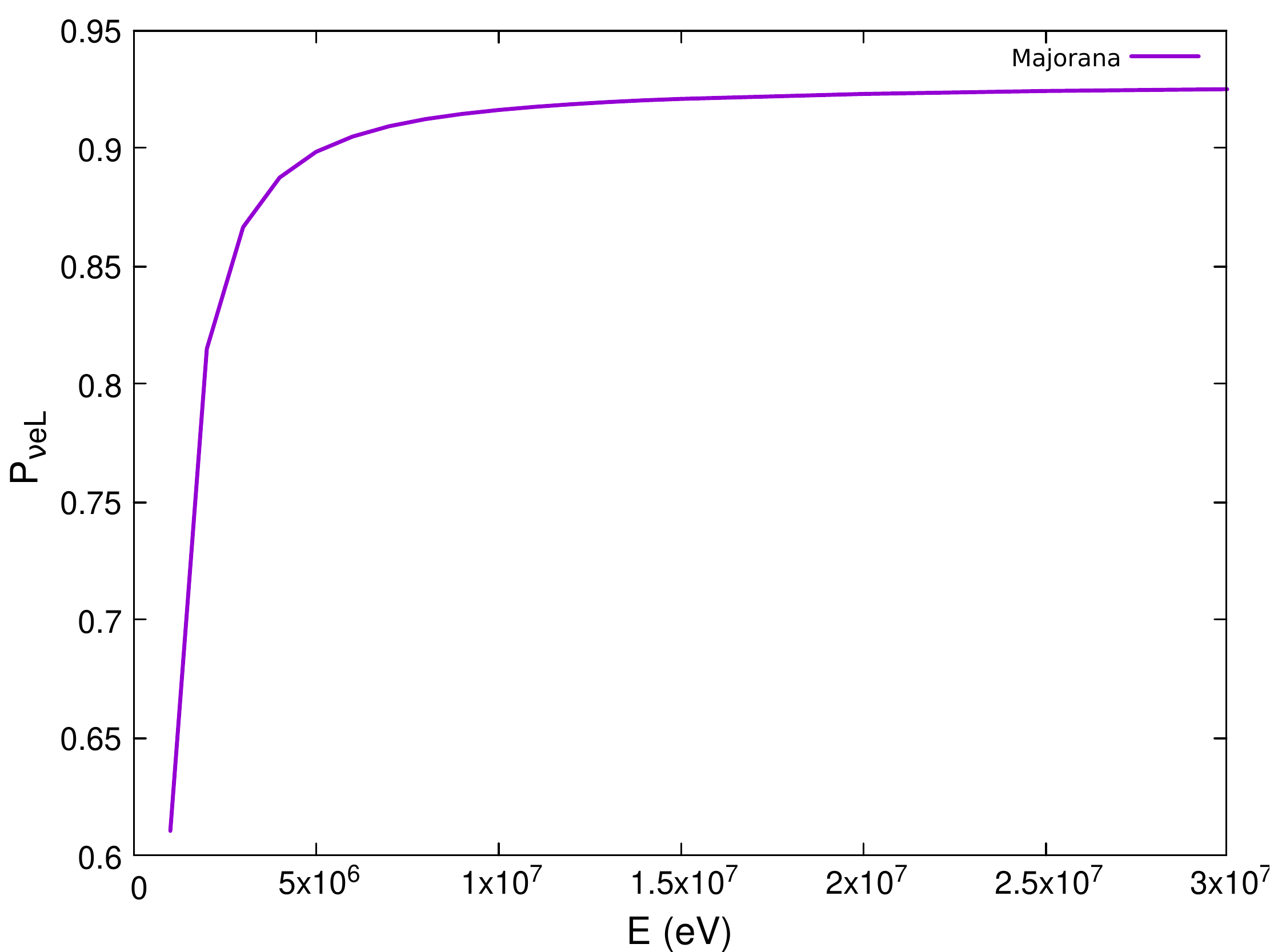}
\caption{Variation of neutrino spin flip survival probability ($P_{\nu_{e L}}$) with energy ($E$) on the surface of the magnetar ($R=11.76$ km). The left (right) panel depicts the Dirac (Majorana) scenario.}
\end{figure}\label{f2}

In fig. 2, we display the variation of survival probability with the neutrino energy, as estimated on the surface of the magnetar. It is deluding from the two panels of the figure that the survival probability becomes nearly identical for the neutrinos with higher values of energies ($E\gtrsim 5$ MeV). It is  also apparent that at $E\sim 1$ MeV, the survival probability is $\sim 0.65$ for Dirac neutrinos, while it is $\sim 0.6$ for Majorana neutrinos. Further,  the difference between the Dirac and Majorana neutrinos are well apprehensible for  neutrinos with lower energies. At higher energies, $E\geq 5$ MeV, the survival probability is saturated to a constant value of $\sim 0.9$ for both scenarios.

\begin{figure}
\includegraphics[scale=0.4]{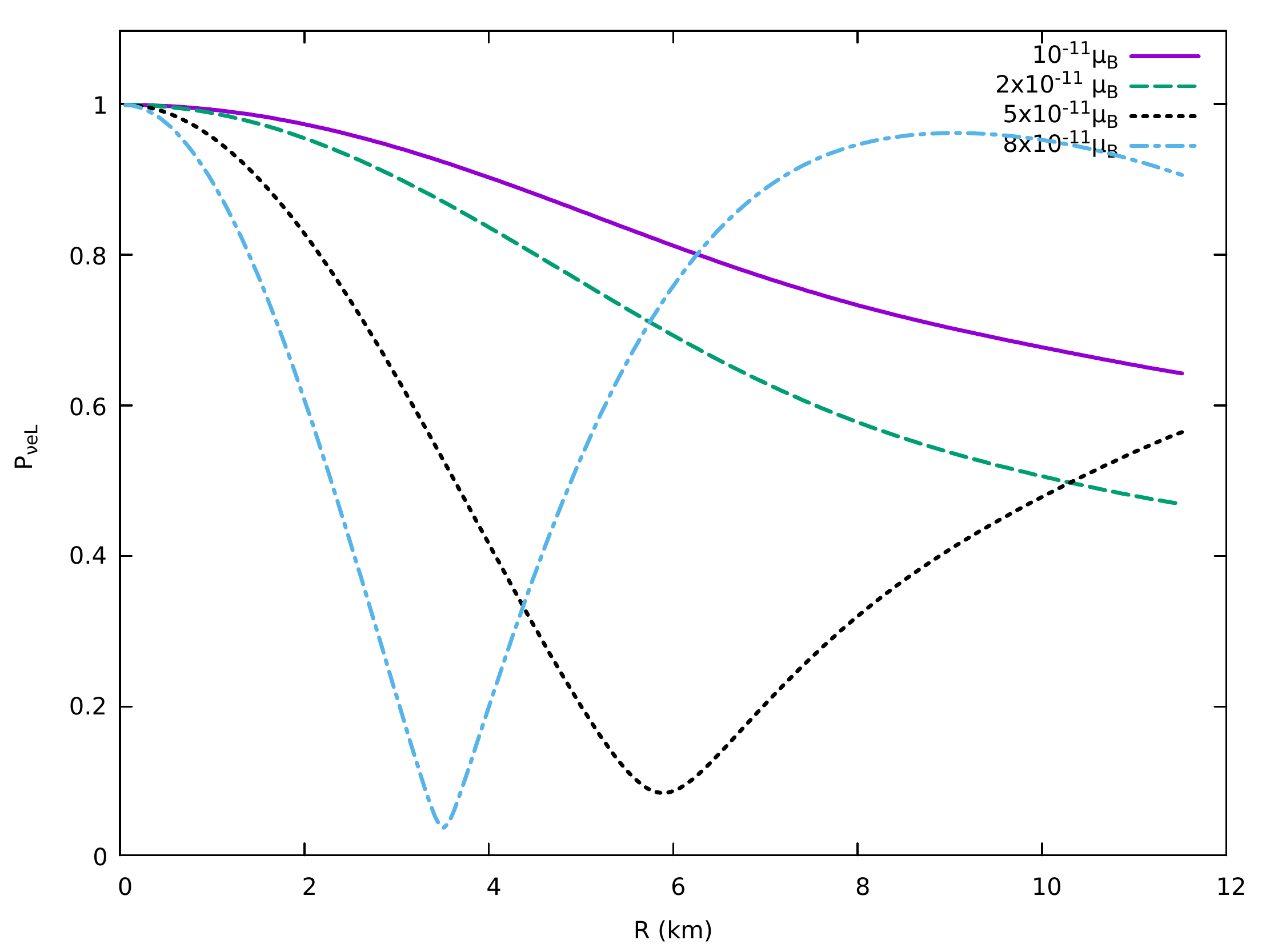}\hspace*{1mm}\includegraphics[scale=0.4]{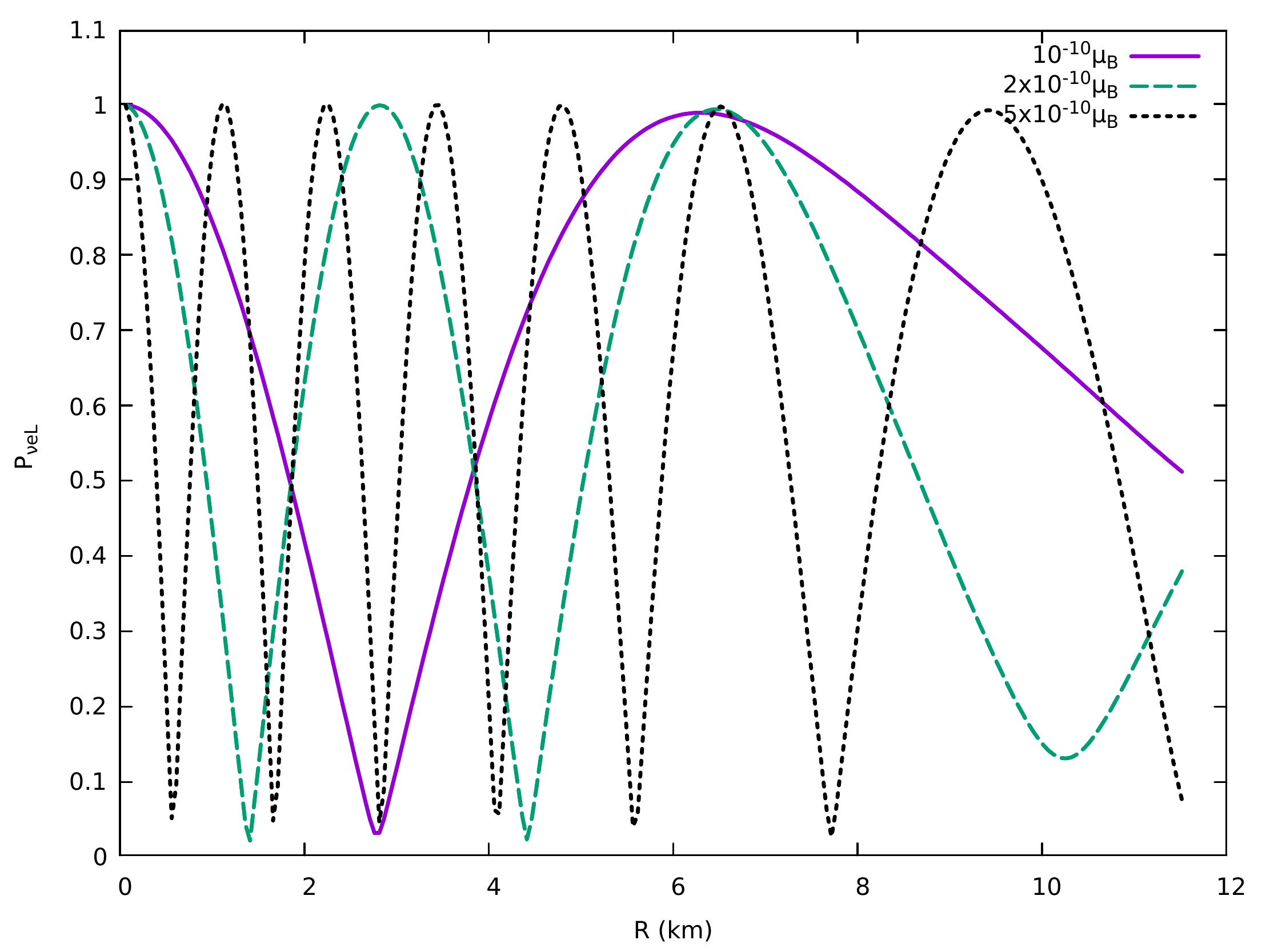}
\caption{Variation of survival probability of $\nu_{eL}$ for Dirac neutrinos with the radial distance ($R$) from the center of the star  for different magnetic moment  ($\mu_{\nu}$) values.}
\end{figure}\label{f3}

In fig. 3 we perform a sensitivity test of neutrino survival probability with the value of the neutrino magnetic moment in the case of Dirac neutrinos. It can be observed from the plots that the spin flip survival probability is extremely sensitive to the value of magnetic moment. It is also avouched that the oscillatory nature of the probability is increasing gradually with higher magnetic moments. In the case of Majorana neutrinos, we arrive at similar conclusions.

\section{Conclusion}\label{conc}
In this work, we study the impingement of the magnetic field of a neutron star on the propagation of neutrinos produced internally under the assumption that the neutrino possesses a finite magnetic moment owing to quantum loop contributions. We consider an isolated magnetar with mass $2.3 M_{\odot}$ and radius $11.76$ km with the central magnetic field of order $10^{18}$ G and the surface field $\sim 10^{15}$ G. The interior matter is governed by a nonlinear Walecka model with GM1 parametrization. The value of magnetic moment  used in our analysis is obtained from the measurements of various terrestrial detectors and different astrophysical and cosmological estimations. The interaction of the neutrino magnetic moment with the magnetic field of the neutron star induces the phenomenon of spin flip oscillations. This results in the reduction of neutrino flux which is $\sim 5\%$ larger for Majorana neutrinos as compared to Dirac at the surface of the neutron star which is favorable from the experimental perspective. We find that the distinction between the Dirac and Majorana nature of neutrinos is visible to a larger extent near the surface region of the star for low energy neutrinos coming from the inner core of the magnetar towards its surface. At higher energies, the distinction between the two kinds of neutrinos is nearly negligible.  Further, the survival probability is extremely sensitive to the assumed value of the neutrino magnetic moment for both Dirac and Majorana neutrinos. Also, the oscillatory nature of the survival probability seems to be more perceptible for larger values of the magnetic moment. Overall, it can be concluded from our analysis that a magnetar is a very suitable system with its extreme interior matter conditions and high magnetic field for the analysis of the phenomenon of spin flavour precession and to discriminate between  the Dirac and Majorana nature of neutrinos.


\begin{thebibliography}{10}

\bibitem{Giunti:2007ry}
C.~Giunti and C.~W.~Kim,
``Fundamentals of Neutrino Physics and Astrophysics,''

\bibitem{Jamieson:2022uya}
B.~Jamieson [Hyper-Kamiokande],
``Future Neutrino Experiments,''
[arXiv:2207.05044 [hep-ex]].

\bibitem{Giunti:2014ixa}
C.~Giunti and A.~Studenikin,
``Neutrino electromagnetic interactions: a window to new physics,''
Rev. Mod. Phys. \textbf{87}, 531 (2015)
[arXiv:1403.6344 [hep-ph]].

\bibitem{Shrock:1982sc}
R.~E.~Shrock,
``Electromagnetic Properties and Decays of Dirac and Majorana Neutrinos in a General Class of Gauge Theories,''
Nucl. Phys. B \textbf{206}, 359-379 (1982)

\bibitem{Fukugita:1987ti}
M.~Fukugita and T.~Yanagida,
``A Particle Physics Model for Voloshin-Vysotskii-Okun Solution to the Solar Neutrino Problem,''
Phys. Rev. Lett. \textbf{58}, 1807 (1987)

\bibitem{Beda:2012zz}
A.~G.~Beda, V.~B.~Brudanin, V.~G.~Egorov, D.~V.~Medvedev, V.~S.~Pogosov, M.~V.~Shirchenko and A.~S.~Starostin,
``The results of search for the neutrino magnetic moment in GEMMA experiment,''
Adv. High Energy Phys. \textbf{2012}, 350150 (2012)

\bibitem{Borexino:2017fbd}
M.~Agostini \textit{et al.} [Borexino],
``Limiting neutrino magnetic moments with Borexino Phase-II solar neutrino data,''
Phys. Rev. D \textbf{96}, no.9, 091103 (2017)
[arXiv:1707.09355 [hep-ex]].

\bibitem{Miranda:2020kwy}
O.~G.~Miranda, D.~K.~Papoulias, M.~T\'ortola and J.~W.~F.~Valle,
``XENON1T signal from transition neutrino magnetic moments,''
Phys. Lett. B \textbf{808}, 135685 (2020)
[arXiv:2007.01765 [hep-ph]].

\bibitem{XMASS:2020zke}
K.~Abe \textit{et al.} [XMASS],
``Search for exotic neutrino-electron interactions using solar neutrinos in XMASS-I,''
Phys. Lett. B \textbf{809}, 135741 (2020)
[arXiv:2005.11891 [hep-ex]].

\bibitem{Raffelt:1990pj}
G.~G.~Raffelt,
``New bound on neutrino dipole moments from globular cluster stars,''
Phys. Rev. Lett. \textbf{64}, 2856-2858 (1990)

\bibitem{Raffelt:1989xu}
G.~G.~Raffelt,
``Core Mass at the Helium Flash From Observations and a New Bound on Neutrino Electromagnetic Properties,''
Astrophys. J. \textbf{365}, 559 (1990)

\bibitem{Raffelt:1992pi}
G.~Raffelt and A.~Weiss,
``Nonstandard neutrino interactions and the evolution of red giants,''
Astron. Astrophys. \textbf{264}, 536-546 (1992)

\bibitem{Fujikawa:1980yx}
K.~Fujikawa and R.~Shrock,
``The Magnetic Moment of a Massive Neutrino and Neutrino Spin Rotation,''
Phys. Rev. Lett. \textbf{45}, 963 (1980)

\bibitem{Pal:1981rm}
P.~B.~Pal and L.~Wolfenstein,
``Radiative Decays of Massive Neutrinos,''
Phys. Rev. D \textbf{25}, 766 (1982)

\bibitem{Voloshin:1986ty}
M.~B.~Voloshin and M.~I.~Vysotsky,
``Neutrino Magnetic Moment and Time Variation of Solar Neutrino Flux,''
Sov. J. Nucl. Phys. \textbf{44}, 544 (1986)
ITEP-1-1986.

\bibitem{Babu:2020ivd}
K.~S.~Babu, S.~Jana and M.~Lindner,
``Large Neutrino Magnetic Moments in the Light of Recent Experiments,''
JHEP \textbf{10}, 040 (2020)
[arXiv:2007.04291 [hep-ph]].

\bibitem{Healey:2013vka}
K.~J.~Healey, A.~A.~Petrov and D.~Zhuridov,
``Nonstandard neutrino interactions and transition magnetic moments,''
Phys. Rev. D \textbf{87}, no.11, 117301 (2013)
[erratum: Phys. Rev. D \textbf{89}, no.5, 059904 (2014)]
[arXiv:1305.0584 [hep-ph]].


\bibitem{Papoulias:2015iga}
D.~K.~Papoulias and T.~S.~Kosmas,
``Neutrino transition magnetic moments within the non-standard neutrino\textendash{}nucleus interactions,''
Phys. Lett. B \textbf{747}, 454-459 (2015)
[arXiv:1506.05406 [hep-ph]].

\bibitem{Kharlanov:2020cti}
O.~G.~Kharlanov and P.~I.~Shustov,
``Effects of nonstandard neutrino self-interactions and magnetic moment on collective Majorana neutrino oscillations,''
Phys. Rev. D \textbf{103}, no.9, 095004 (2021)
[arXiv:2010.05329 [hep-ph]].




\bibitem{Aboubrahim:2013yfa}
A.~Aboubrahim, T.~Ibrahim, A.~Itani and P.~Nath,
``Large Neutrino Magnetic Dipole Moments in MSSM Extensions,''
Phys. Rev. D \textbf{89}, no.5, 055009 (2014)
[arXiv:1312.2505 [hep-ph]].

\bibitem{Fukuyama:2003uz}
T.~Fukuyama, T.~Kikuchi and N.~Okada,
``Neutrino magnetic moments and minimal supersymmetric SO(10) model,''
Int. J. Mod. Phys. A \textbf{19}, 4825-4834 (2004)
[arXiv:hep-ph/0306025 [hep-ph]].

\bibitem{Giunti:2008ve}
C.~Giunti and A.~Studenikin,
``Neutrino electromagnetic properties,''
Phys. Atom. Nucl. \textbf{72}, 2089-2125 (2009)
[arXiv:0812.3646 [hep-ph]].

\bibitem{Xing:2012gd}
Z.~z.~Xing and Y.~L.~Zhou,
``Enhanced Electromagnetic Transition Dipole Moments and Radiative Decays of Massive Neutrinos due to the Seesaw-induced Non-unitary Effects,''
Phys. Lett. B \textbf{715}, 178-182 (2012)
[arXiv:1201.2543 [hep-ph]].

\bibitem{Beda:2009kx}
A.~G.~Beda, E.~V.~Demidova, A.~S.~Starostin, V.~B.~Brudanin, V.~G.~Egorov, D.~V.~Medvedev, M.~V.~Shirchenko and T.~Vylov,
``GEMMA experiment: Three years of the search for the neutrino magnetic moment,''
Phys. Part. Nucl. Lett. \textbf{7}, 406-409 (2010)
[arXiv:0906.1926 [hep-ex]].

\bibitem{TEXONO:2006xds}
H.~T.~Wong \textit{et al.} [TEXONO],
``A Search of Neutrino Magnetic Moments with a High-Purity Germanium Detector at the Kuo-Sheng Nuclear Power Station,''
Phys. Rev. D \textbf{75}, 012001 (2007)
[arXiv:hep-ex/0605006 [hep-ex]].

\bibitem{Derbin:1993wy}
A.~I.~Derbin, A.~V.~Chernyi, L.~A.~Popeko, V.~N.~Muratova, G.~A.~Shishkina and S.~I.~Bakhlanov,
``Experiment on anti-neutrino scattering by electrons at a reactor of the Rovno nuclear power plant,''
JETP Lett. \textbf{57}, 768-772 (1993)

\bibitem{Allen:1992qe}
R.~C.~Allen, H.~H.~Chen, P.~J.~Doe, R.~Hausammann, W.~P.~Lee, X.~Q.~Lu, H.~J.~Mahler, M.~E.~Potter, K.~C.~Wang and T.~J.~Bowles, \textit{et al.}
``Study of electron-neutrino electron elastic scattering at LAMPF,''
Phys. Rev. D \textbf{47}, 11-28 (1993)

\bibitem{LSND:2001akn}
L.~B.~Auerbach \textit{et al.} [LSND],
``Measurement of electron - neutrino - electron elastic scattering,''
Phys. Rev. D \textbf{63}, 112001 (2001)
[arXiv:hep-ex/0101039 [hep-ex]].


\bibitem{Heger:2008er}
A.~Heger, A.~Friedland, M.~Giannotti and V.~Cirigliano,
``The Impact of Neutrino Magnetic Moments on the Evolution of Massive Stars,''
Astrophys. J. \textbf{696}, 608-619 (2009)
[arXiv:0809.4703 [astro-ph]].

\bibitem{Borisov:2014cqa}
A.~V.~Borisov and P.~E.~Sizin,
``Plasmon decay to a neutrino pair via neutrino electromagnetic moments in a strongly magnetized medium,''
[arXiv:1406.3301 [hep-ph]].

\bibitem{deGouvea:2012hg}
A.~de Gouvea and S.~Shalgar,
``Effect of Transition Magnetic Moments on Collective Supernova Neutrino Oscillations,''
JCAP \textbf{10}, 027 (2012)
[arXiv:1207.0516 [astro-ph.HE]].

\bibitem{Vassh:2015yza}
N.~Vassh, E.~Grohs, A.~B.~Balantekin and G.~M.~Fuller,
``Majorana Neutrino Magnetic Moment and Neutrino Decoupling in Big Bang Nucleosynthesis,''
Phys. Rev. D \textbf{92}, no.12, 125020 (2015)
[arXiv:1510.00428 [astro-ph.CO]].

\bibitem{Viaux:2013hca}
N.~Viaux, M.~Catelan, P.~B.~Stetson, G.~Raffelt, J.~Redondo, A.~A.~R.~Valcarce and A.~Weiss,
``Particle-physics constraints from the globular cluster M5: Neutrino Dipole Moments,''
Astron. Astrophys. \textbf{558}, A12 (2013)
[arXiv:1308.4627 [astro-ph.SR]].

\bibitem{Broggini:2012df}
C.~Broggini, C.~Giunti and A.~Studenikin,
``Electromagnetic Properties of Neutrinos,''
Adv. High Energy Phys. \textbf{2012}, 459526 (2012)
[arXiv:1207.3980 [hep-ph]].

\bibitem{Joshi:2019dcj}
S.~Joshi and S.~R.~Jain,
``Neutrino spin-flavor oscillations in solar environment,''
[arXiv:1906.09475 [hep-ph]].

\bibitem{Yilmaz:2016ilw}
D.~Yilmaz,
``Combined effect of NSI and SFP on solar electron neutrino oscillation,''
Adv. High Energy Phys. \textbf{2016}, 1435191 (2016)
[arXiv:1601.03161 [hep-ph]].

\bibitem{Lichkunov:2020zzx}
A.~Lichkunov, A.~Popov and A.~Studenikin,
``Neutrino eigenstates and flavour, spin and spin-flavour oscillations in a constant magnetic field,''
[arXiv:2012.06880 [hep-ph]].

\bibitem{PhysRevD.104.023018}
H.~Sasaki and T.~Takiwaki  
Phys. Rev. D \textbf{104} 023018 (2021)
Adv. High Energy Phys. \textbf{2012}, 459526 (2012)
  
\bibitem{Adhikary:2022phm}
J.~Adhikary, A.~K.~Alok, A.~Mandal, T.~Sarkar and S.~Sharma,
``Neutrino spin-flavour precession in magnetized white dwarf,''
[arXiv:2207.09485 [hep-ph]].

\bibitem{Alok:2022pdn}
A.~K.~Alok, N.~R.~S.~Chundawat and A.~Mandal,
``Cosmic neutrino flux and spin flavor oscillations in intergalactic medium,''
[arXiv:2207.13034 [hep-ph]].
  
  
\bibitem{Peng:2007uu}
Q.~H.~Peng and H.~Tong,
``The physics of strong magnetic fields in neutron stars,''
Mon. Not. Roy. Astron. Soc. \textbf{378}, 159 (2007)
[arXiv:0706.0060 [astro-ph]].

\bibitem{Yakovlev:2000jp}
D.~G.~Yakovlev, A.~D.~Kaminker, O.~Y.~Gnedin and P.~Haensel,
``Neutrino emission from neutron stars,''
Phys. Rept. \textbf{354}, 1 (2001)
[arXiv:astro-ph/0012122 [astro-ph]].

\bibitem{Chrimes:2021wqi}
A.~A.~Chrimes, A.~J.~Levan, P.~J.~Groot, J.~D.~Lyman and G.~Nelemans,
``The Galactic neutron star population \textendash{} I. An extragalactic view of the Milky Way and the implications for fast radio bursts,''
Mon. Not. Roy. Astron. Soc. \textbf{508}, no.2, 1929-1946 (2021)
[arXiv:2105.04549 [astro-ph.GA]].

\bibitem{Kaspi:2017fwg}
V.~M.~Kaspi and A.~Beloborodov,
``Magnetars,''
Ann. Rev. Astron. Astrophys. \textbf{55}, 261-301 (2017)
[arXiv:1703.00068 [astro-ph.HE]].


\bibitem{Menezes:2021jmw}
D.~P.~Menezes,
``A Neutron Star Is Born,''
Universe \textbf{7}, no.8, 267 (2021)
[arXiv:2106.09515 [astro-ph.HE]].


\bibitem{Vidana:2020jhf}
I.~Vida\~na,
``Short introduction to the physics of neutron stars,''
EPJ Web Conf. \textbf{227}, 01018 (2020)

\bibitem{Lattimer:2015eaa}
J.~M.~Lattimer,
``Introduction to neutron stars,''
AIP Conf. Proc. \textbf{1645}, no.1, 61-78 (2015)

\bibitem{Yakovlev:2004yr}
D.~G.~Yakovlev, O.~Y.~Gnedin, M.~E.~Gusakov, A.~D.~Kaminker, K.~P.~Levenfish and A.~Y.~Potekhin,
``Neutron star cooling,''
Nucl. Phys. A \textbf{752}, 590-599 (2005)
[arXiv:astro-ph/0409751 [astro-ph]].


\bibitem{Tsuruta:1986qt}
S.~Tsuruta,
``NEUTRON STAR COOLING,''
Conf. Proc. C \textbf{861214}, 499-503 (1986)

\bibitem{Pethick:1991mk}
C.~J.~Pethick,
``Cooling of neutron stars,''
Rev. Mod. Phys. \textbf{64}, 1133-1140 (1992)

\bibitem{Spruit:2007bt}
H.~C.~Spruit,
``Origin of neutron star magnetic fields,''
AIP Conf. Proc. \textbf{983}, no.1, 391-398 (2008)
[arXiv:0711.3650 [astro-ph]].

\bibitem{Reisenegger:2003pj}
A.~Reisenegger,
``Origin and evolution of neutron star magnetic fields,''
[arXiv:astro-ph/0307133 [astro-ph]].

\bibitem{PhysRevLett.79.2176}
D.~Bandyopadhyay, S.~Chakraborty and S.~ Pal 
Phys. Rev. Lett. \textbf{79} 2176 (1997)  
  
\bibitem{Chatterjee:2018prm}
D.~Chatterjee, J.~Novak and M.~Oertel,
``Magnetic field distribution in magnetars,''
Phys. Rev. C \textbf{99}, no.5, 055811 (2019)
[arXiv:1808.01778 [nucl-th]].
  
\bibitem{Chatterjee:2021wsr}
D.~Chatterjee, J.~Novak and M.~Oertel,
``Structure of ultra-magnetised neutron stars,''
Eur. Phys. J. A \textbf{57}, no.8, 249 (2021)
[arXiv:2108.13733 [nucl-th]].
  

\bibitem{Santos:2004js}
A.~M.~S.~Santos and D.~P.~Menezes,
``Dense and hot matter within the nonlinear Walecka model,''
Phys. Rev. C \textbf{69}, 045803 (2004)

\bibitem{Mueller:1996pm}
H.~Mueller and B.~D.~Serot,
``Relativistic mean field theory and the high density nuclear equation of state,''
Nucl. Phys. A \textbf{606}, 508-537 (1996)
[arXiv:nucl-th/9603037 [nucl-th]].

\bibitem{Glendenning:1991es}
N.~K.~Glendenning and S.~A.~Moszkowski,
``Reconciliation of neutron star masses and binding of the lambda in hypernuclei,''
Phys. Rev. Lett. \textbf{67}, 2414-2417 (1991)
  

\end{thebibliography}
\end{document}